\documentclass[conference]{IEEEtran}
\IEEEoverridecommandlockouts

\usepackage{cite}
\usepackage{amsmath,amssymb,amsfonts}
\usepackage{algorithmic}
\usepackage{graphicx}
\usepackage{textcomp}
\usepackage{xcolor}
\usepackage{booktabs}
\usepackage{multirow}
\usepackage{multicol}

\newcommand{\figOne}{
\begin{figure}[t]
    \centering
    \includegraphics[width=\linewidth]{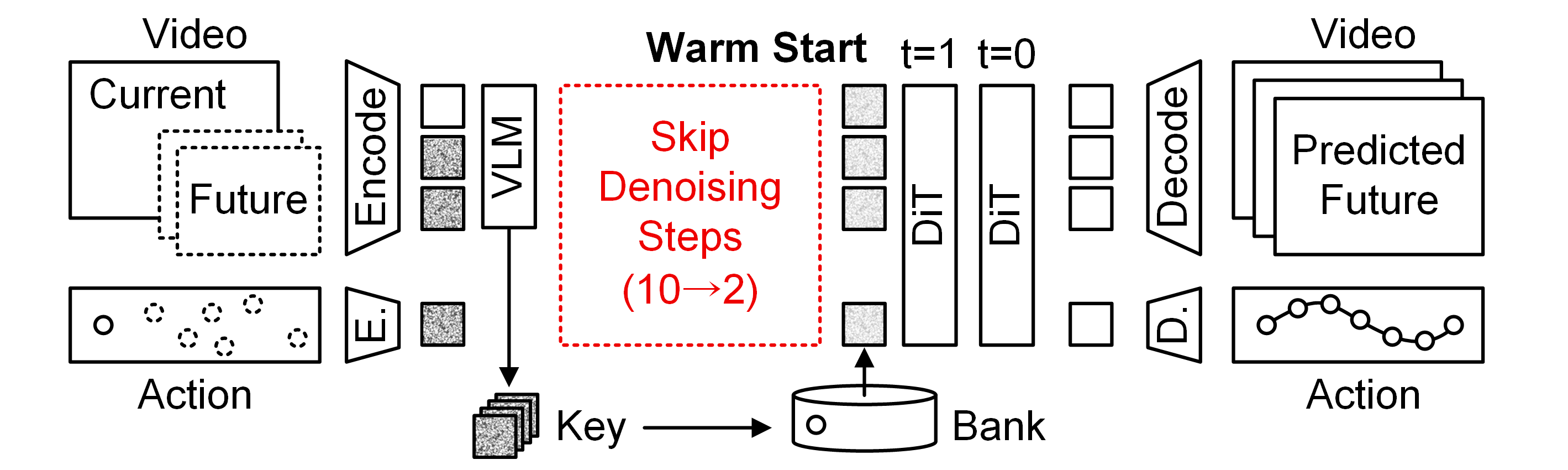}
    \caption{Proposed experience retrieval-based WAM inference using VLM-based retrieval and warm-starting.}
    \label{fig:figOne}
    \vspace{-0.5cm}
\end{figure}
}

\newcommand{\figTwo}{
\begin{figure}[t]
    \centering
    \includegraphics[width=\linewidth]{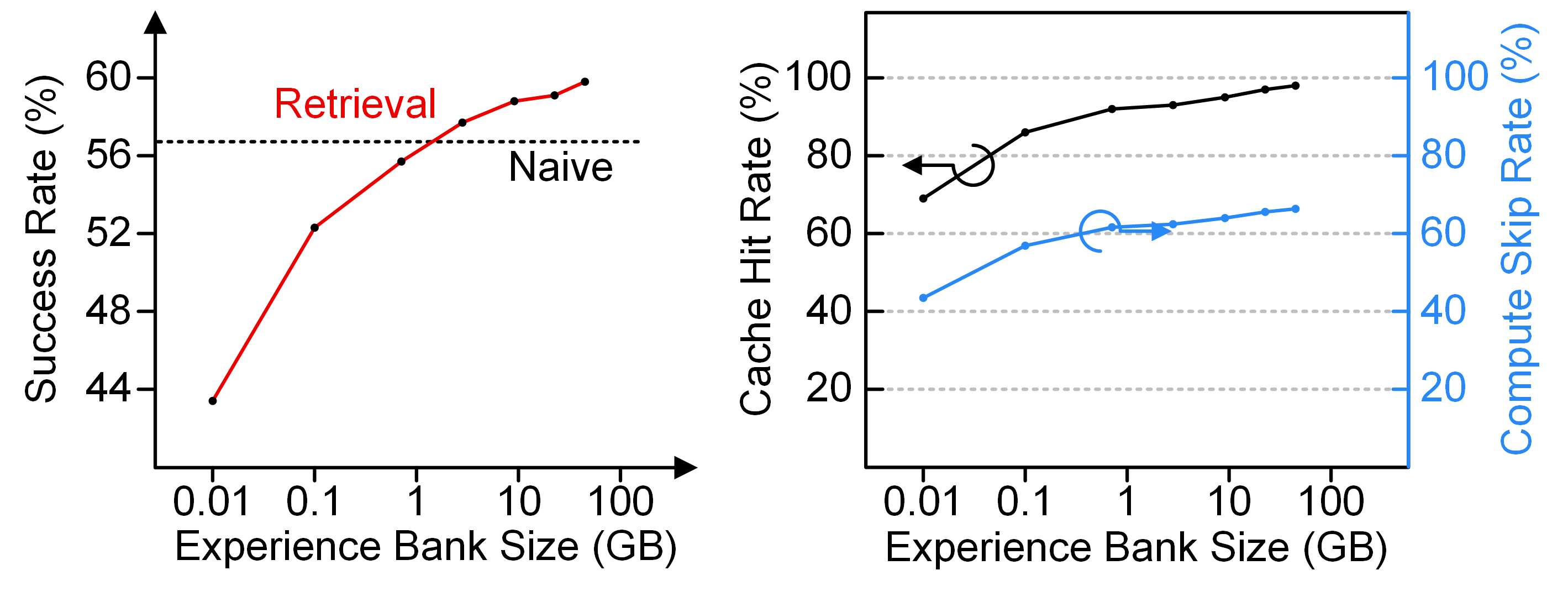}
    \caption{Impact of experience bank scaling on OpenWAM-$\alpha$ in VLABench: (a) task success rate, and (b) retrieval hit rate and computation skipping.}
    \label{fig:figTwo}
    \vspace{-0.5cm}
\end{figure}
}

\newcommand{\figThree}{
\begin{figure}[t]
    \centering
    \includegraphics[width=\linewidth]{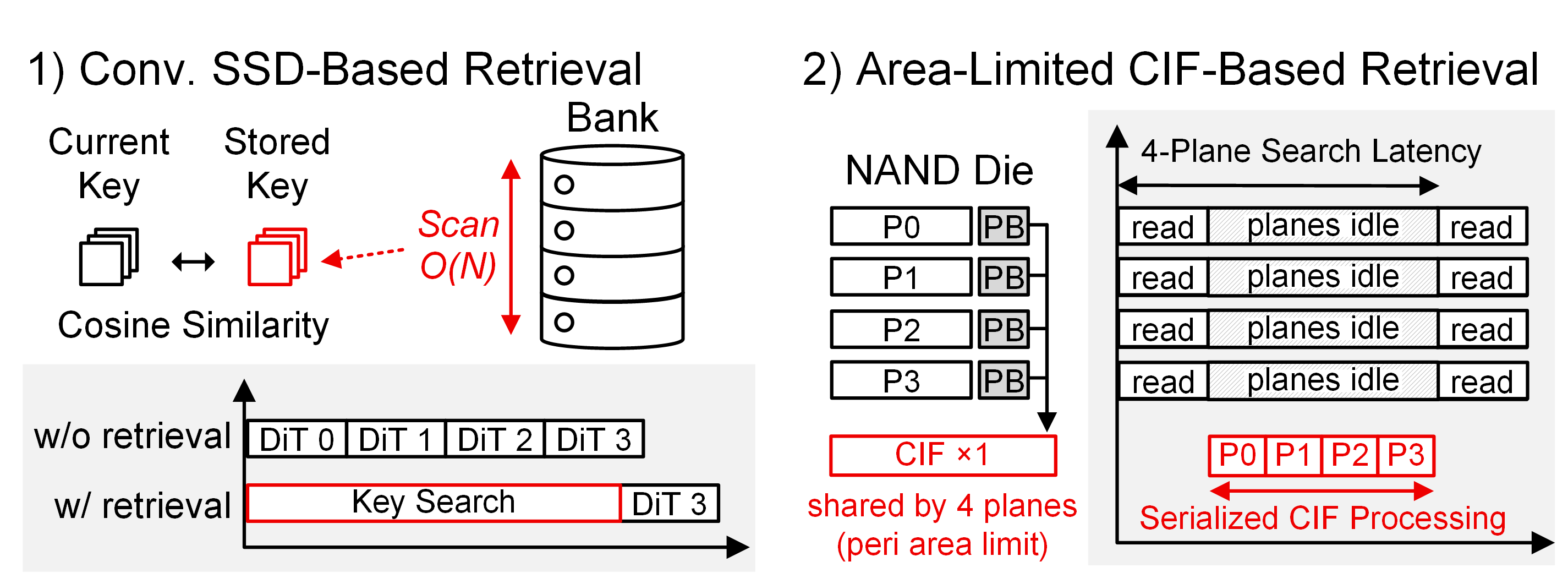}
    \caption{Bottlenecks of conventional experience retrieval on Jetson AGX Orin: (a) SSD-based retrieval over PCIe and (b) CIF-enabled SSD retrieval based on ICE and AiF.}
    \label{fig:figThree}
    \vspace{-0.5cm}
\end{figure}
}

\newcommand{\figFour}{
\begin{figure}[t]
    \centering
    \includegraphics[width=\linewidth]{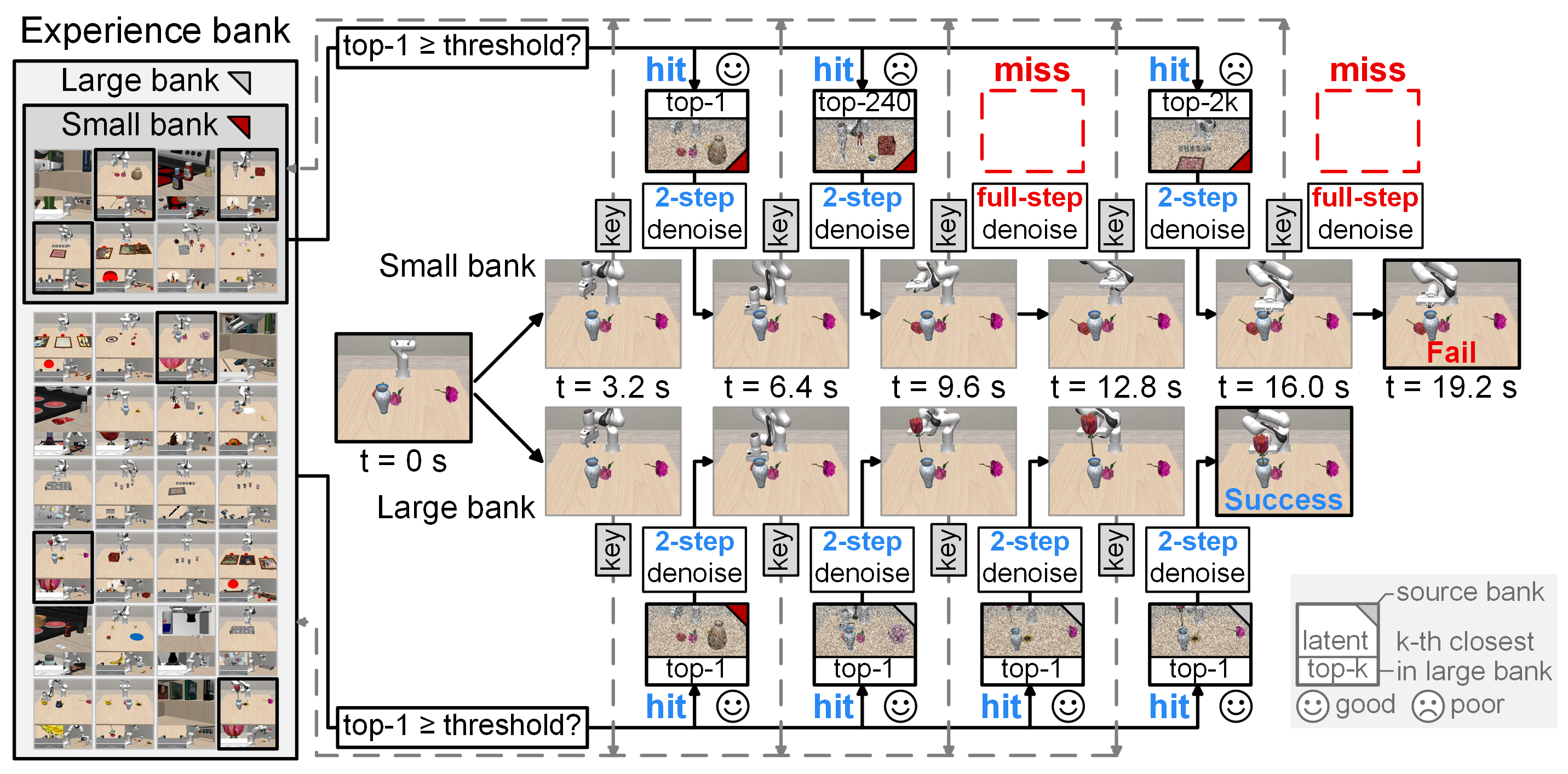}
    \caption{Impact of experience bank size on robotic task execution using successful-experience warm-starting.}
    \label{fig:figFour}
    \vspace{-0.5cm}
\end{figure}
}

\newcommand{\figFive}{
\begin{figure}[t]
    \centering
    \includegraphics[width=\linewidth]{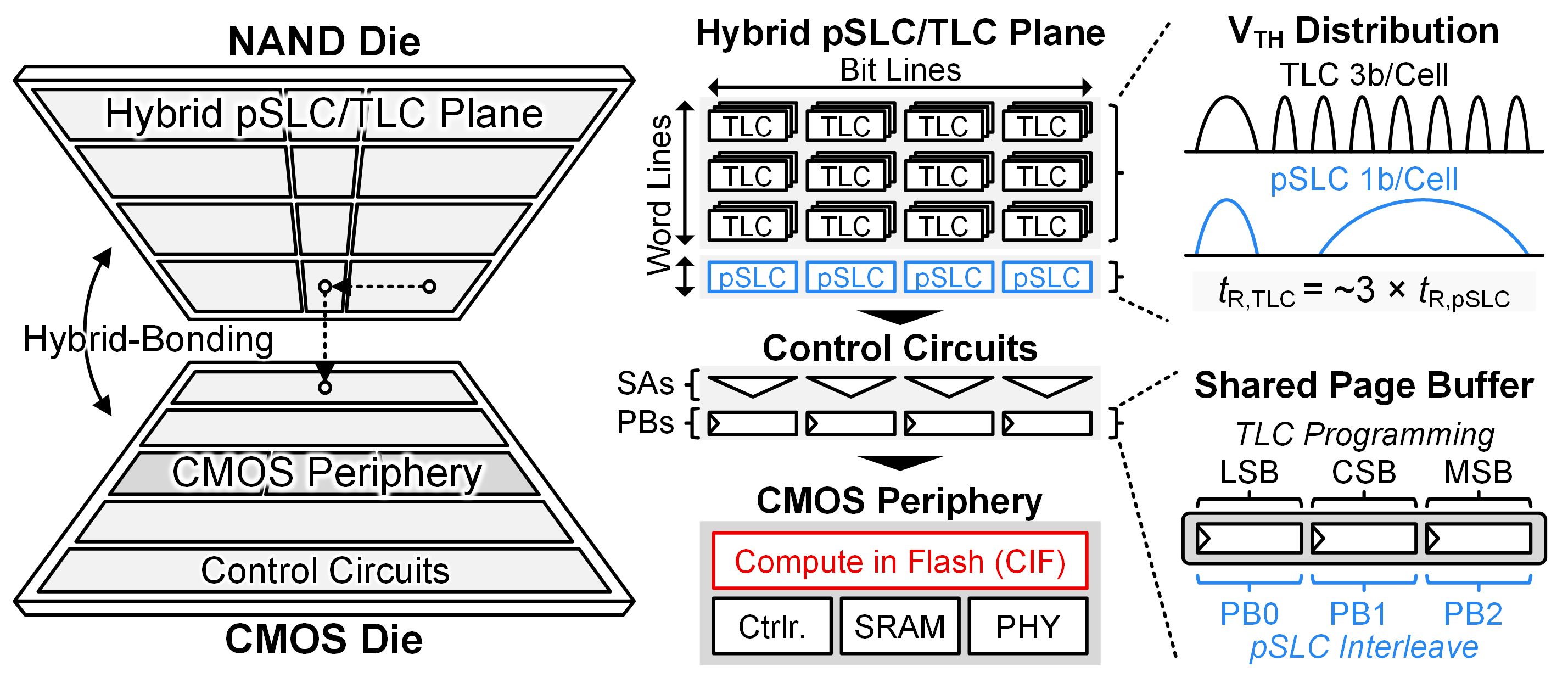}
    \caption{CBA-based FlashX architecture with hybrid pSLC/TLC organization and shared page-buffer reuse.}
    \label{fig:figFive}
    \vspace{-0.5cm}
\end{figure}
}

\newcommand{\figSix}{
\begin{figure}[t]
    \centering
    \includegraphics[width=\linewidth]{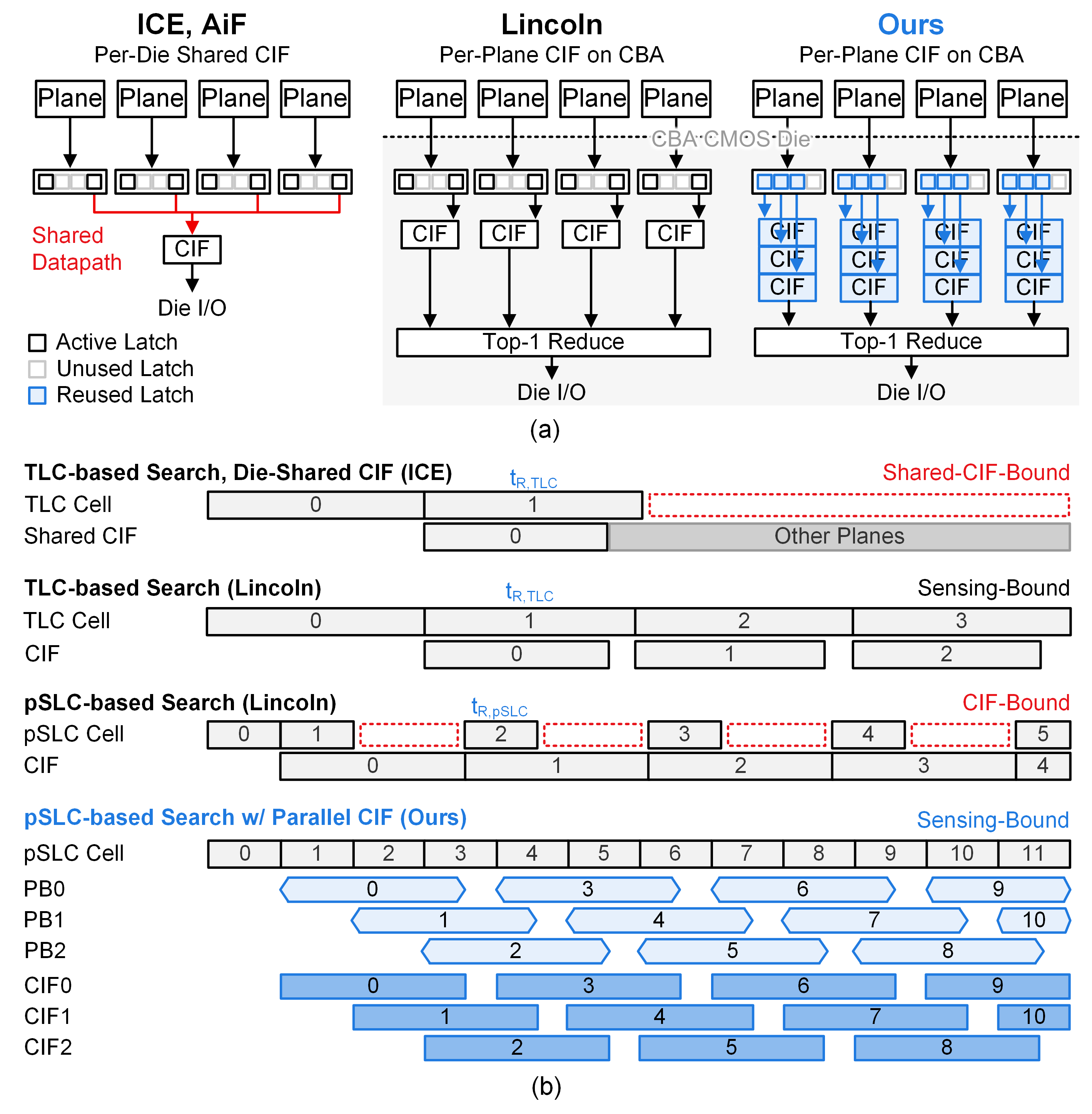}
    \caption{Comparison of conventional and proposed CIF search architectures: (a) hardware organization and (b) search pipeline behavior.}
    \label{fig:figSix}
    \vspace{-0.5cm}
\end{figure}
}

\newcommand{\figSeven}{
\begin{figure}[t]
    \centering
    \includegraphics[width=\linewidth]{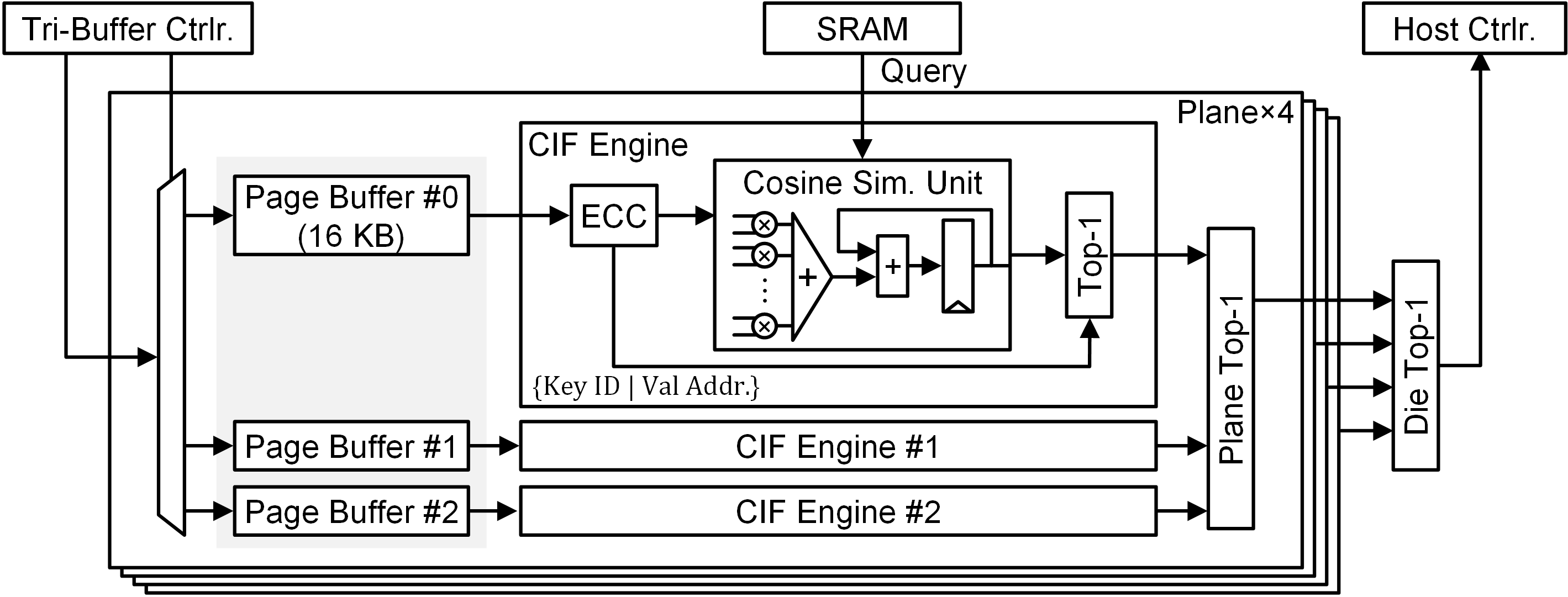}
    \caption{FlashX CIF microarchitecture with interleaved page buffers, parallel similarity engines, and hierarchical top-1 reduction.}
    \label{fig:figSeven}
    \vspace{-0.5cm}
\end{figure}
}

\newcommand{\figEight}{
\begin{figure}[t]
    \centering
    \includegraphics[width=\linewidth]{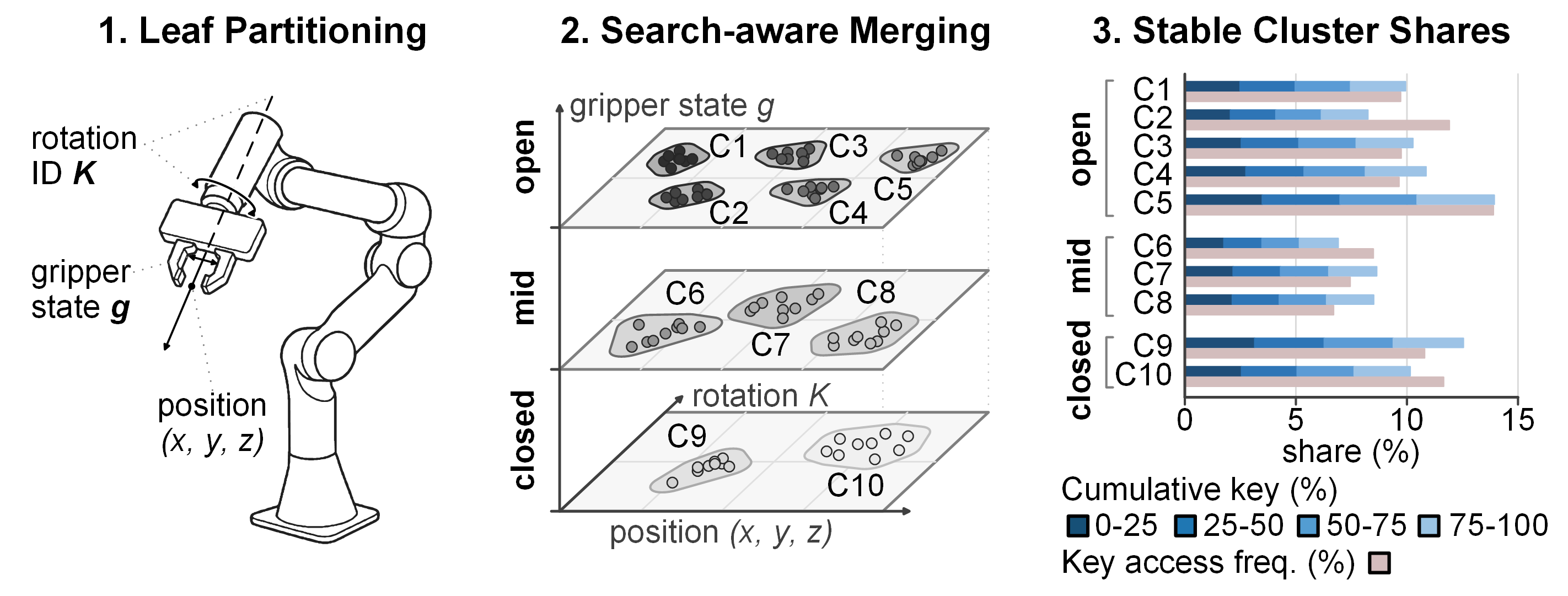}
    \caption{Proprioception-aware experience clustering for selective search and efficient online placement using $\pi_{0.5}$ on LIBERO-Plus.}
    \label{fig:figEight}
    \vspace{-0.5cm}
\end{figure}
}

\newcommand{\figNine}{
\begin{figure}[t]
    \centering
    \includegraphics[width=\linewidth]{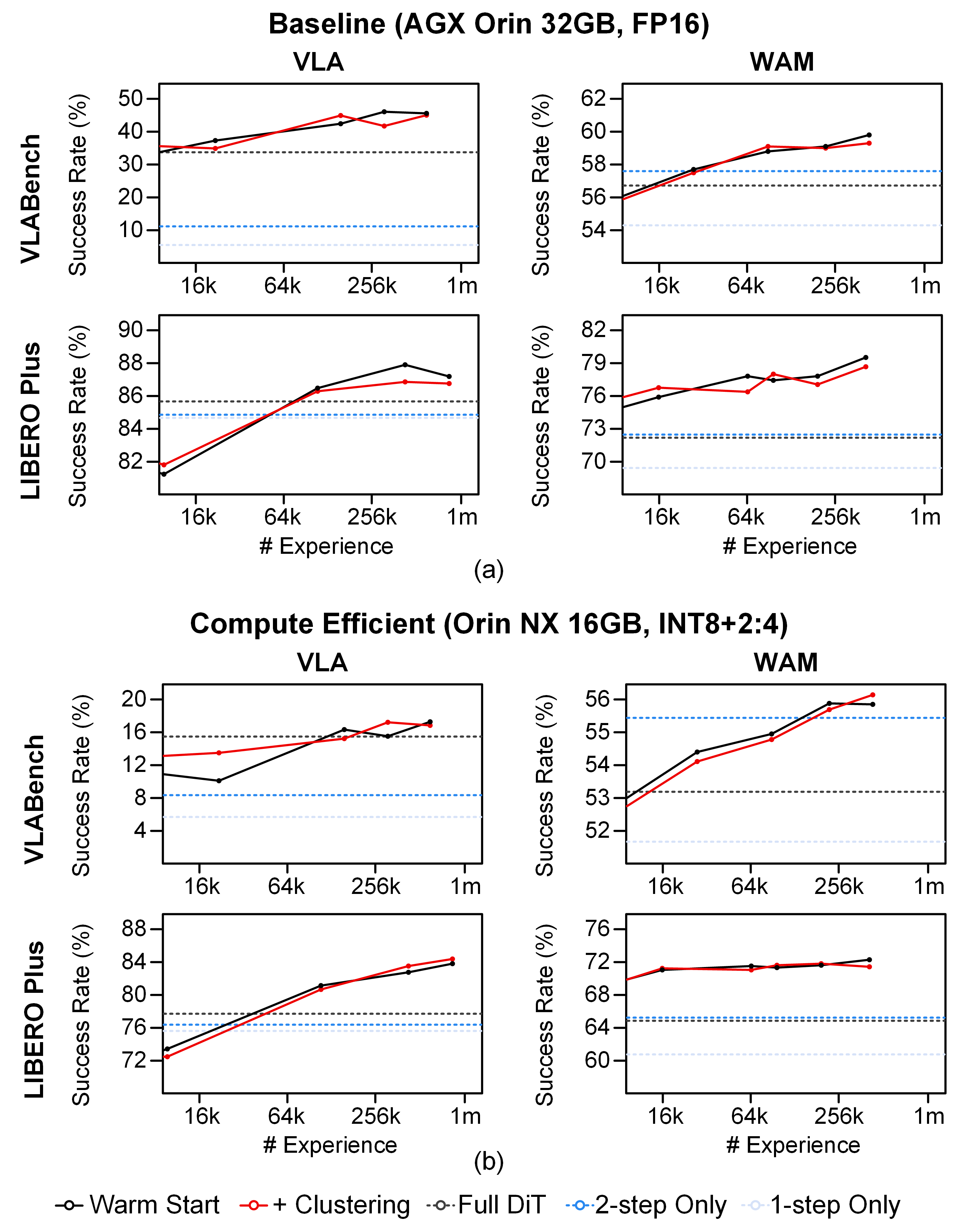}
    \caption{Task success rate scaling under (a) FP16 baseline and (b) INT8 2:4 compute-efficient configurations.}
    \label{fig:figNine}
    \vspace{-0.5cm}
\end{figure}
}

\newcommand{\figTen}{
\begin{figure*}[t]
    \centering
    \includegraphics[width=\linewidth]{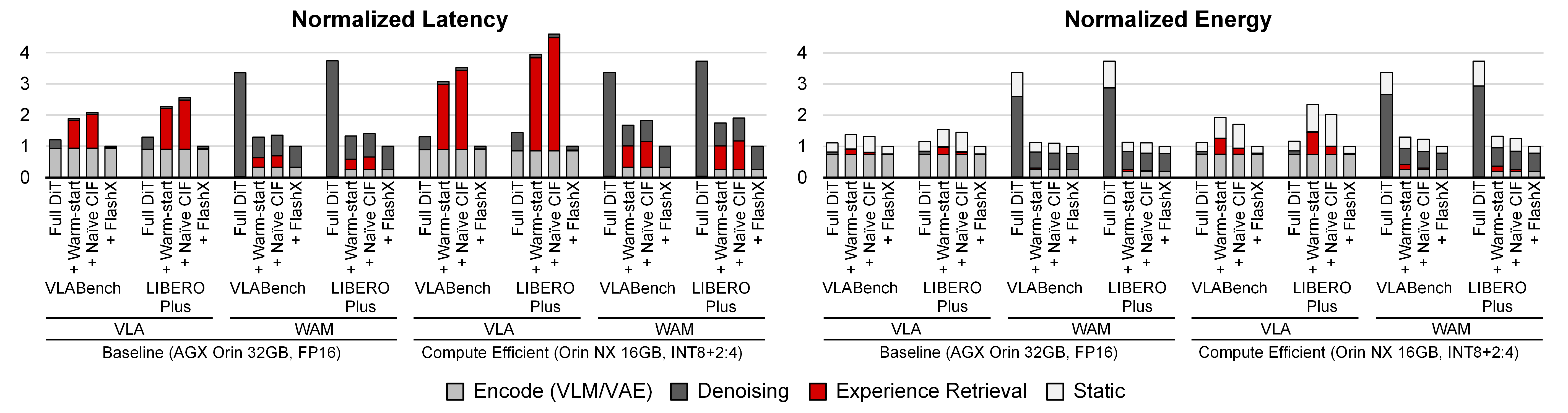}
    \caption{End-to-end inference latency and energy across VLA and WAM workloads under baseline and compute-efficient edge configurations.}
    \label{fig:figTen}
    \vspace{-0.5cm}
\end{figure*}
}

\newcommand{\figEleven}{
\begin{figure}[t]
    \centering
    \includegraphics[width=\linewidth]{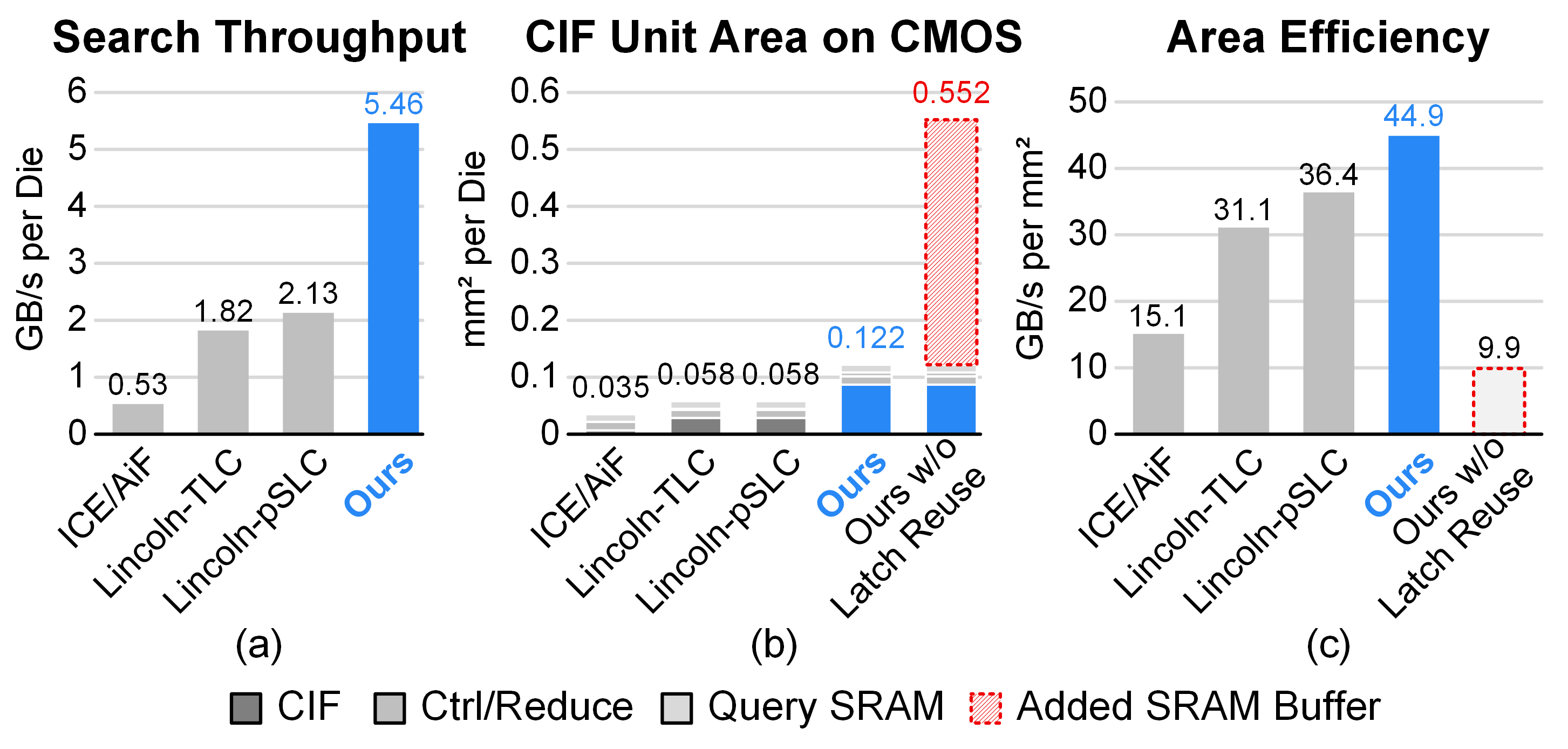}
    \caption{Comparison of NAND search architectures in (a) search throughput, (b) functional-unit area, and (c) throughput per unit area.}
    \label{fig:figEleven}
    \vspace{-0.5cm}
\end{figure}
}

\newcommand{\tabOne}{
\begin{table}[t]
\centering
\footnotesize
\setlength{\tabcolsep}{2pt}
\renewcommand{\arraystretch}{1.12}
\caption{Hardware configuration of FlashX.}
\label{tab:tabOne}
\begin{tabular*}{\linewidth}{@{\extracolsep{\fill}}llcc}
\toprule
\textbf{Component} &
\textbf{Parameter} &
\textbf{AGX Orin} &
\textbf{Orin NX} \\
\midrule
\multirow{3}{*}{xPU}
& GPU & 1792 CUDA & 1024 CUDA \\
& FP16 Dense (TFLOPS) & 27 & 13 \\
& INT8 2:4 (TOPS) & 108 & 50 \\
\midrule
\multirow{3}{*}{Main Memory}
& Type & \multicolumn{2}{c}{LPDDR5} \\
& Capacity (GB) & 32 & 16 \\
& Bandwidth (GB/s) & 204.8 (x256) & 102.4 (x128) \\
\midrule
\multirow{7}{*}{Storage}
& Type & \multicolumn{2}{c}{CBA-based 3D NAND (TLC)} \\
& Capacity (GB) & \multicolumn{2}{c}{2,048 (16 $\times$ 1 Tb)} \\
& Bandwidth (GB/s) & \multicolumn{2}{c}{7.88 (PCIe Gen4 x4)} \\
& Organization & \multicolumn{2}{c}{8 ch, 2 dies/ch, 4 planes/die} \\
& NAND I/O & \multicolumn{2}{c}{x8, 4.8 Gb/s/pin} \\
& Page Size (KB) & \multicolumn{2}{c}{16} \\
& Read Latency ($\mu$s) & \multicolumn{2}{c}{pSLC: 12 / TLC: 36} \\
\bottomrule
\end{tabular*}
\vspace{-2mm}
\end{table}
}

\newcommand{\tabTwo}{
\begin{table}[t]
\centering
\footnotesize
\setlength{\tabcolsep}{2pt}
\renewcommand{\arraystretch}{1.12}
\caption{Per-die hardware overhead of FlashX (45\,nm, 4 planes).}
\label{tab:tabTwo}
\begin{tabular*}{\linewidth}{@{\extracolsep{\fill}}llcc}
\toprule
\textbf{Component} &
\textbf{Implementation} &
\textbf{Area} &
\textbf{Storage} \\
& &
\textbf{(mm$^2$)} &
\textbf{(KB)} \\
\midrule
CIF Engine ($\times$12) & FP16 MAC, 200\,MHz & 0.087 & -- \\
Page Buffer & Reused LSB/CSB/MSB Latches & 0 & (192) \\
Query Buffer & SRAM & 0.013 & 4 \\
Search Ctrl. & Plane/Die Scheduling & 0.014 & -- \\
Top-1 Reduction & Comparator Tree & 0.008 & -- \\
ECC & BCH (10\,b / 1\,KiB) & 0.142 & -- \\
\midrule
\textbf{Total} & & \textbf{0.264} & \textbf{4} \\
\bottomrule
\end{tabular*}
\vspace{-2mm}
\end{table}
}

\def\BibTeX{{\rm B\kern-.05em{\sc i\kern-.025em b}\kern-.08em
    T\kern-.1667em\lower.7ex\hbox{E}\kern-.125emX}}
\begin{document}
\bstctlcite{IEEEexample:BSTcontrol}

\title{FlashX: Scaling Robotic Experience Reuse for Embodied Agents with CBA-Based NAND Flash}


\author{
\IEEEauthorblockN{
Yuseon Choi,
Sanghyeon Park,
Eunjong Lee,
Hanseo Lee,
Sangjin Kim
}
\IEEEauthorblockA{
Gwangju Institute of Science and Technology (GIST), Republic of Korea
}
\IEEEauthorblockA{
yuseon.choi@gist.ac.kr,
\{shpark2623,adria1202,gkstj319\}@gm.gist.ac.kr,
sangjinkim@gist.ac.kr
}
}

\maketitle

\begin{abstract}
Embodied agents are emerging as a promising paradigm for complex robotic tasks, leveraging Vision-Language-Action (VLA) and World Action Models (WAMs) to generate actions through Diffusion Transformer (DiT)-based iterative denoising. However, existing models often struggle to generalize to out-of-distribution (OOD) environments. We observe that retrieving successful experiences and using them to warm-start denoising enables agents to better adapt to unfamiliar situations while reducing computational overhead. This creates a new opportunity to improve task accuracy and inference efficiency simultaneously, in contrast to conventional acceleration techniques that trade accuracy for computation. However, realizing this opportunity requires scaling experience banks, introducing substantial storage and search overhead that challenges the capacity, bandwidth, and power constraints of edge systems.

We propose FlashX, a CMOS Directly Bonded to Array (CBA)-based NAND flash architecture for scalable experience reuse in embodied agents. FlashX exploits CBA's high internal bandwidth for parallel pSLC key search, employing interleaved Compute-in-Flash (CIF) engines to sustain sensing throughput while retaining dense TLC latent storage. Proprioception-aware clustering further reduces search overhead and enables incremental experience acquisition without global reclustering. Evaluations across VLA and WAM workloads demonstrate up to 12.24\%p and 7.43\%p improvements in task success under dense and compute-efficient configurations, respectively. FlashX achieves 2.57$\times$ higher search throughput than the strongest prior NAND search architecture under pSLC sensing, with a further 9.7$\times$ search reduction through clustering. This delivers 1.29--3.95$\times$ end-to-end speedup and 1.12--2.34$\times$ energy efficiency improvement over SSD-based retrieval.

\end{abstract}

\section{Introduction}

Embodied agents have rapidly advanced with the emergence of Vision-Language-Action (VLA) and World Action Models (WAMs)~\cite{CoRL25_OpenVLA,CoRL_pi0.5,arxiv26_openwam}, enabling robots to understand complex environments and execute long-horizon tasks. In particular, Diffusion Transformer (DiT)-based VLA and WAMs~\cite{ICCV23_dit} generate actions and future states through iterative denoising and state prediction. However, existing models often struggle with out-of-distribution (OOD) generalization, as highlighted by VLABench~\cite{ICCV25_vlabench} and LIBERO-Plus~\cite{CVPR26_liberoplus}. Meanwhile, iterative denoising incurs substantial computational overhead and energy consumption. To mitigate these costs, prior studies have explored computation skipping techniques, such as reducing denoising steps~\cite{arxiv26_SnapFlow,RSS24_consistency} and reusing intermediate results~\cite{NeurlIPS25_vlacache,ICRA25_streamingdiffpolicy}. However, these techniques often introduce a trade-off between inference efficiency and task accuracy.

Building on experience-based inference~\cite{Humanoids25_rtcache,arxiv26_actioncache}, we develop a unified warm-start framework for VLA and WAMs that retrieves successful experiences to initialize iterative denoising. Rather than directly reusing past outputs, warm-started inference adapts retrieved latent states to current observations, improving task success while reducing redundant computation. Unlike conventional computation skipping, this approach introduces a new design opportunity that trades storage capacity for both task accuracy and inference efficiency. As the experience bank grows, the likelihood of retrieving relevant successful experiences increases, improving retrieval hit rates and enabling greater computation skipping. More importantly, even after retrieval hit rates saturate, larger banks provide better-matched latent states that further improve task accuracy. 

However, realizing these benefits in edge environments requires a memory architecture capable of efficiently storing and searching large-scale experience banks. While LPDDR provides low access latency, its limited capacity and power budget constrain the retention of continuously growing experiences~\cite{DAC26_slicemoe}. NAND flash offers high storage density and low cost per bit, but large-scale similarity-based key searches incur substantial latency and bandwidth overhead from repeated NAND accesses and data movement~\cite{NeurIPS19_diskann}. Existing Compute-in-Flash (CIF) architectures mitigate data movement by performing search operations within NAND~\cite{mirco22_ice, ISCA25_aif}, yet their search throughput remains constrained by sensing latency and limited computational parallelism. Furthermore, as robotic agents continuously accumulate new experiences, efficient physical organization becomes essential to maintaining search parallelism and minimizing unnecessary key accesses.

In this work, we propose FlashX, a CMOS Directly Bonded to Array (CBA)-based NAND flash architecture for scalable experience reuse in embodied agents. By directly integrating CMOS logic with NAND arrays, CBA enables high-bandwidth internal data access and in-flash computation, reducing search-related data movement. FlashX leverages low-latency pSLC regions for parallel key search, supported by interleaved CIF engines on the CBA CMOS die, while retaining dense TLC regions for latent value storage. Furthermore, FlashX introduces proprioception-aware experience clustering to reduce the search space through selective key retrieval. Online experience placement further eliminates the overhead of frequent global reclustering while maintaining efficient experience organization. Together, these techniques enable capacity-scalable experience reuse, improving task accuracy and inference efficiency under edge memory and power constraints.

\section{Background}

\subsection{Robotic Inference and Experience Reuse}
\figOne

Vision-Language-Action (VLA) models generate robotic actions from visual observations and language instructions. Recent VLAs combine Vision-Language Models (VLMs) with diffusion- or flow-matching-based action heads to generate continuous action chunks~\cite{RSS25_pi0}. World Action Models (WAMs) additionally predict future environmental states alongside robotic actions. For instance, OpenWAM-$\alpha$ combines pretrained video generation with an Action Diffusion Transformer (ActionDiT) to jointly generate future visual states and actions~\cite{arxiv26_openwam}.

Experience-based inference reduces this overhead by retrieving previously successful experiences from a key--value bank. Retrieved experiences can initialize intermediate inference states, directly replay previous actions, or correct generated actions. ActionCache warm-starts denoising using retrieved action representations on VLA~\cite{arxiv26_actioncache}, while RT-Cache directly replays successful action snippets~\cite{Humanoids25_rtcache}. RTCF instead aligns retrieved trajectories with ongoing execution and applies residual corrections to generated actions~\cite{arxiv26_rtcf}. 

\subsection{CBA-Based NAND Flash}

NAND flash provides high-density nonvolatile storage through multilevel cell programming. Triple-Level Cell (TLC) stores three bits per cell, offering high storage density at the cost of multiple sensing operations and increased read latency. In contrast, pseudo-Single-Level Cell (pSLC) operates multilevel NAND cells in a single-bit mode, reducing sensing latency at the expense of storage density.

CMOS Directly Bonded to Array (CBA) technology vertically integrates the NAND cell array and peripheral CMOS circuits through direct bonding~\cite{IEDM23_cba}. This integration provides dense interconnections between the array and CMOS die, supporting high-bandwidth internal data access and logic integration.~\cite{IEDM23_cba, HPCA25_lincoln} Compute-in-Flash (CIF) architectures exploit internal data access to perform computation near stored data, reducing external data movement during similarity search.

\section{Motivation}

\subsection{Opportunities for Experience Scaling}
\figTwo

\figThree

Experience-based inference accumulates successful execution trajectories from long-horizon robotic tasks and reuses them in similar situations. As illustrated in Fig.~1, experiences are collected from successful episodes, while storage and retrieval operate at the granularity of individual action chunks. Unlike conventional computation reuse based on temporal similarity between consecutive frames~\cite{NeurlIPS25_vlacache}, this approach enables experience reuse across temporally distant executions and diverse environments. Retrieved latent states warm-start iterative denoising and are further adapted to current observations, reducing redundant computation while improving task success.

Fig.~2 illustrates two key benefits of experience scaling using OpenWAM-$\alpha$ on VLABench. First, as shown in Fig.~2(a), retrieving successful experiences provides guidance when the original model fails or struggles to determine appropriate actions. Retrieved latent states are adapted to current observations through warm-starting, while larger experience banks provide better-matched experiences that further improve task success. Second, Fig.~2(b) shows that larger experience banks increase retrieval hit rates, enabling more frequent warm-starting. This allows more early denoising steps to be skipped, reducing computational overhead. Consequently, experience scaling simultaneously improves task accuracy and inference efficiency, motivating large-capacity experience storage for embodied agents. However, retaining growing retrieval keys in LPDDR would compete with model execution for limited memory capacity and bandwidth, motivating in-storage search.

\subsection{Challenges of Scalable Experience Retrieval}

However, scaling the experience bank increases retrieval overhead as the footprint of experiences grows to GB scale. Fig.~3(a) illustrates conventional SSD-based retrieval on Jetson AGX Orin~\cite{Nvidia22_jetsonagxorin}, where exhaustive similarity search requires transferring stored keys over PCIe for host-side computation. Repeated NAND accesses and data transfers introduce substantial latency and energy overhead, potentially offsetting the computational savings from warm-started inference. Thus, increasing storage capacity alone is insufficient to achieve efficient experience reuse.

Compute-in-Flash (CIF) architectures reduce external data movement by performing similarity search within NAND. Fig.~3(b) illustrates a CIF-enabled SSD baseline modeled after ICE~\cite{mirco22_ice} and AiF~\cite{ISCA25_aif} on the same Jetson AGX Orin platform. Because the NAND periphery leaves little area for additional logic, a single CIF engine is shared across all planes, serializing similarity computation despite parallel sensing and leaving planes idle. Faster sensing alone cannot relieve this bottleneck, motivating a NAND architecture that pairs low-latency sensing with sufficient in-flash computational parallelism.

\section{FlashX Architecture}

\subsection{Architecture Overview}
\figFour

Fig.~4 illustrates FlashX execution with different experience bank sizes. At each action chunk, the agent generates a VLM-based query to retrieve the top-1 successful experience, using its cached latent state to warm-start denoising when the similarity exceeds a threshold. Otherwise, the agent performs full-step denoising. With a small bank, limited experience coverage leads to frequent misses and less suitable retrieved states. For instance, some top-1 matches from the small bank rank only 240th or 2kth in the larger bank, resulting in suboptimal actions and eventual task failure. In contrast, a larger bank provides more frequent retrieval hits and better-matched latent states, enabling the agent to warm-start inference more consistently and successfully complete the task.

\figFive

Fig.~5 illustrates the CBA-based FlashX architecture, where the NAND array and CMOS periphery are fabricated separately and joined by hybrid bonding, giving each plane a high-bandwidth vertical path to logic on the CMOS die. Each plane stores dense latent values in TLC blocks and retrieval keys in pSLC blocks. Because pSLC separates only two V$_{\text{TH}}$ states with a single sensing step, whereas TLC distinguishes eight states through multiple sensing steps, pSLC reads are about 3$\times$ faster. Key data sensed from the NAND array is transferred through sense amplifiers (SAs) and page buffers (PBs) on the CBA CMOS die to the CIF engines. To keep pace with this faster key supply, FlashX reuses the LSB, CSB, and MSB data latches that every TLC page buffer already holds for programming. Unlike prior work that uses idle latches as static storage for the input vector~\cite{ISCA25_aif}, FlashX rotates sensed key pages through them as three interleaved key buffers (PB0--PB2), each feeding its own CIF engine without additional page-sized storage.

\subsection{Interleaved Parallel CIF Architecture}

\textbf{Baseline CIF Organizations.}
Fig.~6 compares CIF organizations using the same CIF engine. The die-shared designs introduced in Section~III, ICE~\cite{mirco22_ice} and AiF~\cite{ISCA25_aif}, attach a single CIF to all planes of a conventional NAND die. Consequently, similarity computation is serialized across planes, making these designs CIF-bound regardless of sensing latency. Thus, AiF's faster LSB-only sensing provides limited throughput benefits. Lincoln~\cite{HPCA25_lincoln} instead bonds a separate logic layer to the NAND array and places one CIF beneath each plane, sizing it to match the plane's read bandwidth. With TLC reads, this per-plane CIF keeps pace with sensing. When keys are read from pSLC, however, sensing becomes about three times faster while the CIF remains sized for TLC, so the plane sits idle between reads and gains little from pSLC. 

\textbf{Latch-Interleaved CIF Execution.}
FlashX closes this gap by interleaving three CIF engines per plane. Each sensed page is first captured by the sense latch and then moved to one of the three TLC programming-data latches, which holds the page only while its CIF processes it. As shown at the bottom of Fig.~6, page $p_i$ occupies latch and engine $i \bmod 3$, so a latch is released exactly when the page three reads later arrives. As long as the CIF processing time does not exceed three pSLC read times, consecutive pages rotate through the three latches and engines without stalling sensing, and the plane remains sensing-bound at the full pSLC read rate, achieving about three times the search throughput of the single-CIF design. The local top-1 results of the three engines are then merged by plane- and die-level reduction.

\textbf{Design Rationale.}
Interleaving reuses the three TLC programming-data latches as independent buffers, avoiding two additional 16-KB SRAM buffers per plane whose area exceeds that of all CIF logic on the die. FlashX replicates the TLC-sized CIF engine with fixed latch-to-engine connections, eliminating the need for a crossbar. In TLC mode, the latches retain their programming roles, while the cache latch remains available for normal I/O.

\figSix

\subsection{Compute-in-Flash Search Engine}

\textbf{Latch-to-CIF Datapath.} 
Fig.~7 shows the CIF microarchitecture replicated for each of the four planes. In pSLC search mode, a tri-buffer controller steers consecutive sensed key pages into page buffers \#0--\#2, which are the reused LSB, CSB, and MSB data latches, and starts the CIF engine bound to each buffer. 

\textbf{CIF Engine and Top-1 Reduction.} 
Each CIF engine reads one codeword per cycle from its buffer and corrects it through an inline ECC decoder. The cosine similarity unit multiplies the FP16 key elements by the query, which is held in an on-die SRAM and broadcast to all engines, and accumulates the products exactly in fixed point, so the score is independent of processing order. Since keys and queries are L2-normalized, the dot product equals cosine similarity. Each key record carries a header \{Key ID $|$ Value Address\}, which the engine forwards with its running top-1 score; keys with uncorrectable errors are excluded and reported. Plane-level and die-level top-1 units then reduce the engine results, and only the winning key ID and its TLC value address are returned to the host controller, avoiding the transfer of keys outside the NAND.

\section{FlashX Operation}

\subsection{Cluster-Based Experience Mapping}

\textbf{Proprioception-Aware Clustering.}
FlashX exploits the observation that, while visual scenes vary substantially across robotic experiences, the underlying embodiment has a constrained configuration space. As illustrated in Fig.~8, each experience is assigned a base cell address $(g, r, i_x, i_y, i_z)$ from its recorded proprioceptive state. The gripper state is divided into three bins (open, mid, and closed) using fixed thresholds, the end-effector rotation is assigned to the nearest of a fixed set of reference orientations (12 or 24 for the Franka Panda~\cite{ieeeram22_frankapanda}), and the end-effector position is quantized on a fixed 40-cm grid. Base cells are then merged into 9--11 search clusters through initial calibration. Given a query, FlashX searches only the cluster corresponding to its proprioceptive state, retaining the global top-1 match for 93.5\% of queries while scanning approximately 9\% of keys on VLABench.

\subsection{Online Experience Placement}

\textbf{Incremental Experience Acquisition.}
FlashX fixes clustering parameters during calibration, avoiding periodic global reclustering. Successful experiences are assigned to predefined clusters at action-chunk granularity, with unseen cells mapped to their nearest known cell based on proprioception. Keys are distributed across cluster-associated pSLC planes, while latent values are striped across TLC planes. This enables incremental acquisition without relocating existing experiences. Fig.~8 shows that, as the bank grows to 888K keys using $\pi_{0.5}$~\cite{CoRL_pi0.5} on LIBERO-Plus, cluster key shares remain nearly constant, with query access shares closely tracking key distributions.
\figSeven
\figEight

\section{Implementation}
\tabOne

\textbf{Hardware Implementation.}
We implement FlashX's latch-interleaving controller, CIF engines with inline ECC and FP16 cosine similarity, and hierarchical top-1 reduction in RTL. The design is verified against a software golden model and synthesized using Nangate45~\cite{nangate45} at 200 MHz. ICsprout55 design rules are referenced for thick-oxide high-voltage transistors. SRAM is modeled using CACTI~\cite{TACO17_cacti7}, and ECC area is estimated from prior work.

\textbf{NAND and System Modeling.}
We model TLC/pSLC sensing, page-buffer occupancy, and interleaved CIF execution using RTL-measured processing cycles. Table~I summarizes the hardware configuration. TLC read latency is set to 36~$\mu$s based on Kioxia's silicon measurements~\cite{ISSCC25_Kioxia}, while pSLC latency is scaled to 12~$\mu$s following the 1-3-3 Gray-code sensing scheme. All CIF baselines use identical computation units. Search latency is integrated with GPU inference, memory accesses, and PCIe transfers to estimate end-to-end performance on Jetson AGX Orin~\cite{Nvidia22_jetsonagxorin} and Orin NX~\cite{Nvidia22_jetsonorinnx}.

\textbf{Evaluation Framework.}
We evaluate $\pi_{0.5}$ (VLA)~\cite{CoRL_pi0.5} and OpenWAM-$\alpha$ (WAM)~\cite{arxiv26_openwam} on VLABench~\cite{ICCV25_vlabench} and LIBERO-Plus~\cite{CVPR26_liberoplus}, targeting out-of-distribution (OOD) robotic generalization with the Franka Panda~\cite{ieeeram22_frankapanda}. Experience banks contain successful action chunks, using $\pi_{0.5}$ VLM features as retrieval keys and action or video/action latents as values for VLA and WAM, respectively. Both models use FP16 precision and 10-step denoising, with two-step warm-starting when top-1 cosine similarity exceeds a P95-calibrated threshold; otherwise, full inference is performed. We evaluate dense models on AGX Orin~\cite{Nvidia22_jetsonagxorin} and SparseGPT~\cite{ICML23_sparsegpt}-based 2:4 sparsity with weight-only INT8 on Orin NX~\cite{Nvidia22_jetsonorinnx}. We compare full inference, SSD retrieval, CIF retrieval, and FlashX in task success, search throughput, area, and end-to-end latency and energy.
\figNine

\section{Evaluation}
\figTen

\textbf{Success Rate from Experience Scaling.}
Fig.~9 evaluates task success rates as the experience bank scales across VLA and WAM workloads on VLABench~\cite{ICCV25_vlabench} and LIBERO-Plus~\cite{CVPR26_liberoplus}. We compare full-step inference, 1-/2-step denoising without experience reuse, and experience-based two-step warm-starting with exhaustive or cluster-restricted retrieval. As shown in Fig.~9(a), larger experience banks provide better-matched latent states, improving task success by up to 12.24\%p over conventional inference. The benefits extend toward one million experiences, suggesting further opportunities for accuracy improvement with continued experience scaling. In contrast, reducing denoising steps without retrieved experiences can substantially degrade accuracy, highlighting the importance of experience-guided warm-starting. Cluster-restricted retrieval preserves nearly the same task success as exhaustive search while scanning only approximately 10\% of stored keys. Fig.~9(b) further evaluates models with SparseGPT~\cite{ICML23_sparsegpt}-based 2:4 sparsity and weight-only INT8 quantization. Experience scaling improves task success by up to 7.43\%p even when combined with these compute-efficient techniques, demonstrating that experience reuse is orthogonal to model-level computation reduction.

\figEleven

\textbf{End-to-End Performance and Energy Efficiency.}
Fig.~10 compares end-to-end latency and energy across VLA and WAM workloads, normalized to FlashX. Although warm-starting reduces denoising from 10 to 2 steps, SSD-based retrieval incurs substantial search overhead for 1M experiences, often offsetting the computational savings and even slowing down VLA inference. Na\"ive CIF avoids external key transfers but remains bottlenecked by its die-shared computation engine, failing to exploit NAND read parallelism. FlashX addresses both limitations through CBA-based pSLC sensing and parallel CIF execution (13.6$\times$), followed by cluster-restricted search (9.7$\times$), achieving 132$\times$ faster retrieval. Consequently, FlashX achieves 1.29--3.95$\times$ end-to-end speedup over SSD-based warm-starting, enabling computational savings to translate into actual inference acceleration. The benefits become more pronounced on compute-efficient models, where reduced GPU computation increases the relative cost of retrieval. FlashX further improves energy efficiency by 1.12--2.34$\times$ by eliminating external key traffic and reducing system static energy during retrieval.

\textbf{Comparison with Prior NAND Search Architectures.}
Fig. 11 compares per-die search throughput and area efficiency across NAND search architectures using identical FP16 CIF engines at 200~MHz. ICE~\cite{mirco22_ice} and AiF~\cite{ISCA25_aif} share a single CIF across planes, limiting throughput to 0.53~GB/s despite AiF's faster LSB-only sensing. Lincoln~\cite{HPCA25_lincoln} places one CIF per plane, achieving 1.82~GB/s with TLC. RTL simulation shows that processing a 16-KB page takes 6,167 cycles (30.8~$\mu$s), fitting within TLC sensing (36~$\mu$s) but exceeding pSLC sensing (12~$\mu$s) by 2.6$\times$. Thus, Lincoln gains only 1.17$\times$ from pSLC, whereas FlashX interleaves three CIF engines per plane to sustain pSLC sensing throughput, reaching 5.46~GB/s (3.00$\times$ over Lincoln-TLC). Despite the additional engines, FlashX requires only 0.122~mm$^2$ of logic per die, equivalent to 0.34\% of the NAND die area. Reusing existing TLC programming-data latches avoids 0.430~mm$^2$ of additional SRAM buffers, which would increase the logic area by 4.5$\times$. Consequently, FlashX achieves 44.9~GB/s/mm$^2$ area efficiency, outperforming Lincoln-TLC by 1.44$\times$, whereas the SRAM-buffered variant drops to 9.9~GB/s/mm$^2$.

\tabTwo

\textbf{Hardware Analysis.}
Table~II summarizes FlashX's per-die CMOS area. Logic is synthesized with Nangate~\cite{nangate45} 45\,nm, and SRAM is estimated using CACTI~\cite{TACO17_cacti7}. Each CIF meets the 200-MHz target with a 1.99-ns critical path. Three CIFs per plane are required to match pSLC sensing, as a single CIF takes 2.6$\times$ longer than a pSLC read. Given pSLC's lower raw bit error rate, we model lightweight BCH error correction based on prior work~\cite{ISCA25_aif}. FlashX exploits this integration to couple parallel CIF engines with existing LSB/CSB/MSB programming latches, sustaining pSLC sensing throughput without the 0.43\,mm$^2$ of additional SRAM buffers. Overall, FlashX adds 0.264\,mm$^2$ of logic and ECC per die, occupying only 0.75\% of a 35\,mm$^2$ NAND die.

\section{Discussion}
FlashX's benefits depend on experience reuse opportunities and the scale of the experience bank. With sufficient successful experiences, retrieved latent states can improve task success while reducing denoising computation. However, these benefits may be limited when relevant experiences are scarce or current observations differ substantially from stored experiences. Furthermore, proprioception-based clustering exploits the fixed embodiment of a robot, requiring initial calibration when applied to different robotic platforms. This work focuses on continuous experience accumulation within a fixed embodiment, leaving cross-embodiment experience sharing and long-term NAND endurance and management optimization for future work.

\section{Conclusion}

This paper presents FlashX, a CBA-based NAND architecture that improves task success and inference efficiency through large-scale experience reuse for embodied agents. FlashX combines parallel pSLC key search with dense TLC latent storage and addresses in-flash search bottlenecks through parallel CIF execution. Proprioception-based clustering and incremental experience placement further reduce the search space without global reclustering. Our evaluation demonstrates that FlashX improves task success as experience banks scale, overcomes the throughput limitations of prior NAND search architectures, and translates the computational savings of experience-based warm-starting into end-to-end performance and energy efficiency gains.


\bibliographystyle{IEEEtran}
\bibliography{refs}


\end{document}